\documentstyle[12pt]{article}
\newcommand{\be}{\begin{equation}}
\newcommand{\nn}{\nonumber}
\newcommand{\bea}{\begin{eqnarray}}
\newcommand{\eea}{\end{eqnarray}}
\newcommand{\ba}{\begin{array}}
\newcommand{\ea}{\end{array}}
\newcommand{\ee}{\end{equation}}

\expandafter\ifx\csname mathbbm\endcsname\relax

\else

\fi \textheight 22cm \textwidth 15cm \topmargin 1mm \oddsidemargin
5mm \evensidemargin 5mm

\newcommand{\beas}{\begin{eqnarray*}}
\newcommand{\eeas}{\end{eqnarray*}}
\newcommand{\bes}{\begin{equation*}}
\newcommand{\ees}{\end{equation*}}

\newcommand{\lf}{\left}
\newcommand{\ri}{\right}
\newcommand{\f}{\frac}
\newcommand{\dagg}{\dagger}

\def\tr           {\mbox{\rm tr}\,}

\def\i2           {\mbox{$\frac{i}{2}$}}

\def\al           {\alpha}

\def\bet           {\beta}
\def\beb           {{\bar \beta}}

\def\del           {\delta}

\def\ep           {\epsilon}

\def\et           {\eta}

\def\ga           {\gamma}

\def\la           {\lambda}

\def\om           {\omega}
\def\Om           {\Omega}

\def\vph           {\varphi}
\def\ps           {\psi}

\def\rh           {\rho}

\def\si           {\sigma}

\def\th{\theta}

\def\pl           {\partial}

\def\ran          {\rangle}
\def\lan          {\langle}

\def\Ah{{\hat A}}
\def\ah{{\hat a}}

\def\we {{\wedge}}

\begin{document}

\begin{titlepage}
\hfill \vbox{
    \halign{#\hfil         \cr
           } 
      }  
\vspace*{20mm}
\begin{center}
{\LARGE \bf{{Dual Instantons in Anti-membranes \\

\vspace{2mm}

Theory  }}}\\ 

\vspace*{15mm} \vspace*{1mm} {A. Imaanpur\footnote{aimaanpu@theory.ipm.ac.ir}  
and M. Naghdi\footnote{m.naghdi@modares.ac.ir}}

\vspace*{1cm}

{\it Department of Physics, School of Sciences\\ 
Tarbiat Modares University, P.O. Box 14155-4838, Tehran, Iran}\\

\vspace*{1mm}

\vspace*{1cm}

\end{center}

\begin{abstract}

We introduce two ansatzs for the 3-form potential of Euclidean 11d supergravity 
on skew-whiffed $AdS_4\times S^7$ background which results in two scalar modes with $m^2=-2$ 
on $AdS_4$. Being conformally coupled with a quartic interaction it is possible to find 
the exact solutions of the scalar equation on this background. These modes turn out to be invariant 
under $SU(4)$ subgroup of $SO(8)$ isometry group whereas there are no corresponding $SU(4)$ singlet BPS 
operators of dimensions one or two on the boundary ABJM theory. Noticing the interchange of ${\bf 8}_s$ 
and ${\bf 8}_c$ representations under skew-whiffing in the bulk, we propose the theory of anti-membranes should similarly be obtained from ABJM theory by swapping these representations. In particular, this enables us to 
identify the dual boundary operators of the two scalar modes. We deform the boundary theory by the dual operators 
and examine the fermionic field equations and compare the solutions of the deformed theory with those of the bulk.
 
\end{abstract}


\end{titlepage}

\section{Introduction}

Aharony, Bergman, Jafferis, and Maldacena (ABJM) have recently succeeded to construct a Chern-Simons-matter theory 
which describes the low energy dynamics of $N$ $M2$-branes at the tip of the orbifold 
${\bf C}_4/{\bf Z}_k$ \cite{ABJM}. 
This theory is conjectured to be dual to M-theory on $AdS_4\times S^7/{\bf Z}_k$, 
where $k$ is the level of Chern-Simons on the gauge theory side. 
For large $k$ ($k^5 >>  N$), the dual theory is better described in terms of type IIA string theory on $AdS_4\times CP^3$. In this note, though, we are interested in the $k=1$ case where the boundary theory is conjectured to have an enhanced ${\cal N}=8$ supersymmetry together with a global $SO(8)$ symmetry. 

For the supergravity solution of $AdS_4\times S^7$, one can flip the sign of $F_4$ flux (skew-whiffed) and still get a solution which is maximally supersymmetric. On this background we obtain two supergravity modes with a mass squared $m^2=-2$ which could couple to operators of dimensions $1$ or $2$. Further, with our particular ansatzs these modes turn out to be singlet under $SU(4)$ subgroup of $SO(8)$ isometry group of $S^7$, and hence according to the AdS/CFT 
duality, the dual BPS operators must also be invariant under $SU(4)$. However, in ABJM boundary theory 
there are no such BPS operators of dimension $1$ or $2$ which are $SU(4)$ invariant. 
A look at the supergravity side, though, 
provides a hint; in the skew-whiffed solution of the bulk, which corresponds to anti-membranes, one needs to interchange $8_s$ and $8_c$ representations to get the right amount of supersymmetries upon compactification on the $S^1$ fiber \cite{HOP}. Therefore, we propose the boundary theory of anti-membranes should similarly be related to that of ABJM by swapping these representations. This change of representations, however, is only possible when $k=1$ or $k=2$, where the ABJM Lagrangian 
has an enhanced ${\cal N}=8$ supersymmetry with $SO(8)$ global symmetry. The triality of $SO(8)$ then allows one 
to permute the three 8-dimensional representations, i.e., ${\bf 8}_v$, ${\bf 8}_c$, and  ${\bf 8}_s$, into one another 
and get three inequivalent Lagrangians.  Note that from the standpoint of $SO(8)$ these three Lagrangians are completely equivalent,  but they look different under $SU(4)$ decomposition.    

In ABJM, scalars, fermions and supersymmetry charges 
are decomposed under $SU(4)$ as 
\bea
{\bf 8}_v &=&  {\bf 4} \oplus \bf{ \bar{ 4}} \nn \\
{\bf 8}_c &=&  {\bf 4} \oplus \bf{ \bar{ 4}} \nn \\
{\bf 8}_s &=&  {\bf 6} \oplus {\bf 1} \oplus {\bf 1}\, ,
\eea   
respectively. Hence for anti-membranes, to conform with the supergravity side, fermions should decompose as 
\be
{\bf 8}_s = {\bf 6} \oplus {\bf 1} \oplus {\bf 1}\, ,
\ee
then triality of $SO(8)$ implies that for supercharges we should have    
\be
{\bf 8}_c =  {\bf 4} \oplus \bf{ \bar{ 4}} 
\ee
while scalar decomposition is not changed.  

Now we can see how $SU(4)$ singlet BPS operators of dimension 2 can arise.  
First note that rank two 
symmetric traceless operators of dimension one sit in ${\bf 35}$ representation of $SO(8)$ 
which under $SU(4)$ decomposes as
\be
{\bf 35 }_v =  {\bf 15} \oplus {\bf 10} \oplus {\bf 10}\, .
\ee 
Let $Y^A$ denote four scalar fields in ${\bf 4}$ of $SU(4)$, then the above decomposition corresponds to 
having the following BPS operators of dimension 1:
\bea
O^{A}_{\ B}&=& \tr\lf( Y^A Y^\dagg_B -\f{\del^A_{\ B}}{4}(Y^CY^\dagg_C) \ri)\nn \\
O^{AB}&=& \tr\lf(Y^A T^\dagg  Y^B\ri) \nn \\
\bar {O}_{AB}&=& \tr\lf( Y^\dagg_A T Y^\dagg_B\ri)\, , \nn
\eea
where $T$'s indicate monopole operators (for $k=1$ and $k=2$) which are needed to make gauge invariant operators 
out of two $Y^A$'s \cite{KLE8}.  
Since for anti-membranes supercharges $Q_A$ are in ${\bf \bar {4}}$ of $SU(4)$, we can get singlet scalar operators 
of dimension 2 by acting with supercharges twice on the above operators:
\bea
{\cal O}_1 &=&\{Q_A ,\, [{\bar Q}^B ,\, O^{A}_{\ B}]\}\nn \\
{\cal O}_2 &=& \{Q_A ,\, [Q_B ,\, O^{AB}]\} \nn \\
\bar{{\cal {O}}_2} &=& \{{\bar Q}^A ,\, [{\bar Q}^B ,\, \bar {O}_{AB}]\}\, . \label{DES}
\eea   
These are in fact the three singlets in the decomposition of ${\bf 35 }_s$:
\be
{\bf 35}_s =  {\bf 20} \oplus {\bf 6} \oplus {\bf 6} \oplus {\bf 1} \oplus {\bf 1} \oplus {\bf 1}\, .
\ee 

These three operators have the right symmetry properties to be identified with the three linearized supergravity 
scalar modes that we find in the bulk (note that $SU(4)$ singlet BPS operators of dimension one are still missing, and 
therefore, in the following, we are only considering operators of dimension 2). 
Moreover, neglecting the backreaction, we are able to solve the field equation of $F_4$ exactly. 
Examining the behavior of the solution near the boundary we find that it satisfies a mixed boundary condition. 
A scalar of $m^2=-2$ could couple to operators of dimensions 1 or 2. For a boundary operator with dimension 2 we have to choose the leading term as a source while the subleading term would correspond to the expectation value of that operator. The opposite is true for an operator of dimension one \cite{KLE9}. 
Following \cite{WITMULTI}, we perturb the boundary theory by a dual operator corresponding to ${\cal O}_1$ 
which has no $U(1)_b$ charge. We observe that the new Lagrangian admits exact solutions with no monopole charge, 
and hence they are identified with bulk solutions which are invariant under $SU(4)\times U(1)$.  

Another possibility is to decompose eight scalars as  ${\bf 6} \oplus {\bf 1} \oplus {\bf 1}$ with fermions and 
supercharges in ${\bf 4} \oplus \bf{ \bar{ 4}}$. This time we obtain three scalar BPS operators of dimension one 
which are invariant under $SU(4)$. In contrast to operators in (\ref{DES}), these operators will be primary. 
However, note that in the 11d supergravity the scalars in ${\bf 8}_v$ must decompose as  
${\bf 4} \oplus \bf{ \bar{ 4}}$ to get the 10-dimensional scalars on right representations 
upon compactification \cite{HOP}. Further, the scalar modes that we find in the bulk are all coming from the 3-form 
potential and hence are pseudoscalars whereas the above three scalar operators are real scalars. 
So we conclude that this pattern of decomposition cannot be realized on anti-membranes.  

In section 2 we begin with 11 dimensional skew-whiffed background of $AdS_4\times S^7$. On this background 
we provide two different ansatzs for the 3-form potential reducing the 4-form field equation to a 4d scalar 
equation on $AdS_4$. As the scalar is conformally coupled with a quartic self-interaction we are able to find 
its exact solutions. In section 3 we examine the behavior of this solution near the boundary. Taking the leading 
term as a source we see that the dual operator has to have a dimension 2. We discuss how triality of $SO(8)$ allows 
us to rearrange the field representations in ABJM theory in order to get the anti-membranes theory. 
We deform the Lagrangian with a multi-trace operator and examine the fermionic field equations. 
We then obtain exact solutions when a fermionic field and the $U(1)$ gauge fields are turned on. 
Conclusions and outlook are brought in section 4.        

\section{11d Supergravity in Skew-Whiffed Background}

Let us start with the 11 dimensional supergravity action
\bea
S=\f{-1}{2\kappa_{11}^2}\int d^{11}x\, \sqrt{ g}\, R 
+\f{1}{4\kappa_{11}^2}\int (F_4\wedge *F_4)
+\f{i}{12\kappa_{11}^2}\int A_3\wedge F_4\wedge F_4\, , \label{11}
\eea
where we have included a factor of $i$ in the Chern-Simons term as we work in Euclidean space. 
For the equation of motion of $A_3$ we have
\be
d*F_4=-\f{i}{2}F_4\wedge F_4\, .\label{E11}
\ee

Let us take the background to be $AdS_4\times S^7/{\bf Z}_k$, with the metric
\be
ds^2= \f{R^2}{4}ds^2_{AdS_4} + R^2 ds^2_{S^7/\bf{ Z}_k}\, ,
\ee
where
\be
ds^2_{S^7/\bf{ Z}_k} =\f{1}{k^2}(d\vph +k \omega)^2 + ds^2_{CP^3}\, ,
\ee
and $\om$ is related to the K\"ahler form $J$ of $CP^3$ through
\be
J=d\om \, .
\ee
This metric describes $S^7$ as a $U(1)$ bundle over $CP^3$. $\vph$ parametrizes the $U(1)$ circle with  
radius $R/k$, so for large $k$ the radius of the circle is small and the effective description will be 
in terms of 10d type IIA supergravity. For the background 4-form flux we have 
\be
F_4=\f{3i}{8}R^3\ep_4\, ,\label{SIGN}
\ee
with an $i$ factor in Euclidean space. The sign of 4-form background flux 
is important in getting conformally coupled scalars in 4 dimensions. In fact, the sign in (\ref{SIGN}) 
corresponds to skew-whiffed solutions in Minkowskian signature. To solve the field equations in this background, 
in the following, we consider two different ansatzs which reduce the 4-form field equation to a single scalar 
equation in four dimensions.\footnote{Similar ansatzs have also been independently proposed in \cite{GAUNT1}.} 
Being conformally invariant, we are able to solve the effective 4d equation exactly by neglecting the backreaction. 

Amusingly, the solution will in fact have a zero 4d modified energy momentum tensor and so it will not backreact 
on $AdS_4$ background \cite{DEH, PAPA}. However, it cannot be uplifted to an exact solution in 11 dimensions as the energy momentum tensor does have nonzero components along $S^7$. 
On the other hand, as long as one is only interested in the behavior of the solution near the boundary and the 
correlation functions of the dual operators, one can ignore the backreaction on the metric \cite{SKE}.


\subsection{The First Ansatz}

To write our first ansatz, we note that there are gauge (with respect to the $U(1)$ gauge factor of $CP^3$) 
covariantly constant spinors $\th$, which could be used to construct complex charged 3-forms on $CP^3$ 
\cite{WARN}:
\be
K_{ijk}={\bar \th}\Gamma_{ijk}\th
\ee
such that
\be
dK=4\, i\omega\wedge \, K.
\ee
Now if we define
\be
L=e^{4i\vph/k}\, K
\ee
we have 
\be
dL={4i}(\f{d\vph}{k} +\omega)\we L=4i\,  e^7\we L\label{LL}\, .
\ee
Further, the Hodge dual of $L$ is
\be
*_7L=ie^7\we L
\ee
so that, together with eq. (\ref{LL}), this implies that
\be
dL= 4 *_7L\, .
\ee
The above properties of $L$ allow for the appearance of identical terms on both sides of the equation of 
motion (\ref{E11}) so that we can reduce it to a 4-dimensional equation.

So to proceed, we take the following ansatz for $F_4$:
\be
F_4=N\ep_4 + R^9d(f\, \Om)= N\ep_4 +R^9 df\we \Om + R^9f\, d\Om \, ,\label{ANS0}
\ee
with 
\be
\Om = L+{\bar L}\, ,
\ee
$N$ and $f$ are scalar functions on $AdS_4$, so that $dF_4=0$. We have included an $R^9$ factor to account 
for the dimension 5 of $L$. For the Hodge dual we obtain
\be
*F_4=R^3\lf( \f{8}{3}N J^3\we e^7 -  \f{R^9}{4} *_4df\we *_7\Om + \f{R^9}{16} f\, \ep_4\we *_7d\Om \ri)\, ,\label{STAR}
\ee
where $\ep_{12345\cdots 11}=\ep_{1234}\, \ep_{5\cdots 11}$, and hence the minus sign in the second term.
Note that
\bea
&&*_7\Om =ie^7\we (L-{\bar L}) \nn \\
&& *_7d\Om = *(dL+d{\bar L})=4\Om \, ,\nn
\eea
therefore the equation of motion, (\ref{E11}), reads
\bea
&& \f{8}{3}dN\we J^3\we e^7 - \f{iR^9}{4} d*_4df\we e^7\we (L-{\bar L}) + i R^9f\, 
\ep_4\we e^7\we (L-{\bar L})  \nn \\
&&= -iR^6Nf\ep_4\we d\Om -iR^{15}fdf\we\Om\we d\Om \nn \\
&&= 4R^6Nf \ep_4\we e^7\we (L-{\bar L})+8R^{15} fdf\we e^7 \we L\we{\bar L}\, ,
\eea
where use has been made of
\bea
d*_7\Om=0 \nn \\
d\Om \we d\Om =0 \, .\nn
\eea
Let us normalize $L$ so that
\be
L\we {\bar L}=-\f{i\la }{48R^{10}}\, J^3\label{LL2}
\ee
for some real dimensionless parameter $\la$. Recalling that $L$ is a $(3,0)$ form and 
$J_{\al\beb}=ig_{\al\beb}$, we can see that $\la$ must be positive 
\be
\la > 0\, .
\ee

Using (\ref{LL2}), the equation of motion splits to:
\bea
&& *_4d*_4df - 4 f\,  = 16 i\f{N}{R^3} f \nn \\
&& dN=-i\f{\la}{32}R^5\, df^2 \, .\label{FIR}
\eea
The last equation implies
\be
N=-i\f{\la}{32} R^5 f^2 +\f{3i}{8}R^3\, , \label{N}
\ee
here we have chosen the constant of integration equal to the background field. 
Plugging this back into the first equation of (\ref{FIR}), we have
\be
*d*df +2f - \f{\la R^2}{2} f^3=0\, , \label{LAMBDA}
\ee
with $*$ indicating the Hodge dual on $AdS_4$ with a unit radius. 
Going back to an $AdS_4$ metric with radius $R/2$, we have
\be
\f{1}{\sqrt{g}}\pl_\mu(\sqrt{g}\, \pl^{\mu} f )+\f{8}{R^2}\, f - 2\la\, f^3=0\, ,\label{LAM}
\ee
which we solve at the end of this section.

\subsection{ The Second Ansatz}
For the second ansatz let
\be
F_4=N\ep_4 + R^4 d(f\, \Om)= N\ep_4 +R^4 df\we \Om + R^4 f\, d\Om \, ,\label{SEC}
\ee
where now
\be
\Om = e^7\we J\, ,
\ee
with
\be
d\Om =2*_7\Om\, .
\ee
Repeating steps (\ref{STAR}) to (\ref{N}), we get
\be
*d*df +2f - \f{3R^2}{2} f^3=0\, ,\label{LAMBDAFIX}
\ee
or 
\be
\f{1}{\sqrt{g}}\pl_\mu(\sqrt{g}\, \pl^{\mu} f )+\f{8}{R^2}\, f -{6}\, f^3=0\, ,\label{EEQQ}
\ee
for an $AdS_4$ metric with radius $R/2$. Note that in contrast with our previous ansatz 
where we have an arbitrary ``coupling constant" 
$\la$, here its value gets fixed to $\la =3$. Note that, however, instead of (\ref{N}), we 
have
\be
N=\f{-3i}{16} R^5 f^2 +\f{3i}{8}R^3\, .\label{N1}
\ee
Although, the mass squared $m^2 R^2 =-2$ is negative but 
it is above the lower bound, $-9/4$, for the stability. Further, it falls in the range
\be
-9/4 < m^2 R^2  < -5/4 \, ,
\ee  
permitting a quantization with Dirichlet or Neumann boundary conditions, and hence coupling  
to operators of dimension $2$ or $1$, respectively \cite{KLE9, FREE}. 

Note that if we had chosen 
\be
N_0=-\f{3i}{8}R^3\, ,\label{N2}
\ee
in (\ref{N1}) or (\ref{N}) for the background field (corresponding to 
the Euclidean version of ABJM background), we would have obtained scalars of $m^2R^2=10$, which are not conformally coupled and hence the 4d equations could not be solved exactly. We will comment on the dual operators of these modes in the next section.  

To solve equation (\ref{LAMBDA}) or (\ref{LAMBDAFIX}), we note that the mass term in this equation is such that 
it permits a conformal transformation to the flat space. To see this, let us write the metric of $AdS_4$ in Poincare 
coordinate
\be
ds^2 = \f{R^2}{4z^2}\lf(dz^2 + \del_{ij} dx^i dx^j\ri)\, , \ \ \ i, j=1,2,3
\ee   
so that equation (\ref{LAM}) reads
\be
\f{4z^4}{R^2} \pl_z\lf( \f{\pl_z f}{z^2} \ri) + \f{4z^2}{R^2} \pl_i \pl^i f + \f{8}{R^2} f - 2\la f^3 =0\, . \label{ZZ}
\ee
Now let us make the following scaling transformation on $f$
\be
f=\f{2z}{R}\, g\, .
\ee
In terms of $g$, equation (\ref{ZZ}) simply reduces to the equation of a massless scalar (with the same 
cubic term) on flat space:
\be
\pl_z \pl_z g + \del_{ij} \pl^i \pl^j g  - 2\la g^3 =0\, ,
\ee
which is easily solved for 
\be
g=\f{2}{\sqrt{\la}}\lf(\f{b}{-b^2+(z+a)^2+(\vec{x}-\vec{x_0})^2}\ri)\, ,
\ee
with $b$ a free parameter of the solution. We need to choose $a^2>b^2$ with $a>0$ to have a smooth solution.
So for $f$ we have
\be
f=\f{4}{R\sqrt{\la}}\lf(\f{bz}{-b^2+(z+a)^2+(\vec{x}-\vec{x_0})^2}\ri)\, .\label{SOL}
\ee

Similar equations to (\ref{LAMBDA}) and (\ref{LAMBDAFIX}) have appeared in \cite{DEH, PAPA} and \cite{LORAN}. 
As we noticed above, these 4-dimensional equations are {\em conformally coupled} in the sense that the action 
from which the equations are derived is conformally invariant.
  
It is also possible to make a comparison between our ansatzs and the more general ansatzs that have been independently proposed 
in \cite{GAUNT1}. First note that as we have ignored the back reaction on the metric we should set $U=V=0$ in eq. (2.4) of 
that paper. Further, if we set $H_3=H_2=A_1=h=0$ in eq. (2.5) of that paper and define $f\equiv \chi +{\bar \chi}$ we get our first 
ansatz (\ref{ANS0}). On the other hand, our second ansatz in (\ref{SEC}) is obtained by setting $H_3=H_2=A_1=\chi =0$ with $f\equiv h$ 
in eq. (2.5). For the skew-whiffed solution we see that eq. (B.14) (with $U=V=\chi =0$) of that paper reduces to
\be
f=6(-1 +h^2)
\ee
plugging this back into eq. (B.11) (with $H_3=H_2=A_1=0$) we get 
\be
d*dh + 4h(f+4) {\rm vol}_4  = 0 \, ,
\ee
or
\be
*d*dh +8h -24\, h^3= 0\, ,
\ee
which upon a scaling gives the conformally coupled scalar equation in (\ref{EEQQ}).

\section{Dual Operators}

We would now like to look at the behavior of the solution near the boundary. 
As we will be discussing the second ansatz let us set $\la =3$, and $R=1$ for convenience. 
When $z \to 0$, for solution (\ref{SOL}) we have
\be
f(z,x) \to \al(x)\, z + \bet(x)\, z^2 \, ,\label{AL}
\ee
where
\be
\al(x)= \f{4}{\sqrt{3}}\lf(\f{b}{-b^2+a^2+(\vec{x}-\vec{x_0})^2}\ri)\, \nn
\ee
and
\be
\bet(x)=\f{-16\, ab}{\sqrt{3}\lf({-b^2+a^2+(\vec{x}-\vec{x_0})^2}\ri)^2}\, ,\\ \label{BET}
\ee
are taken to be the source and the expectation value of the dual boundary operator corresponding to the solution 
(\ref{SOL}) in the bulk. As mentioned in the Introduction, since we are taking the leading term in (\ref{AL}) as a source 
the dual operator will have dimension 2. On the other hand, for a bulk mode with $m^2=-2$ it is also possible to take the 
second term as a source and thus the dual operator would have dimension one. In the following, we consider the first 
possibility as there is no $SU(4)$ invariant BPS operator of dimension one in anti-membranes boundary theory. 

We will assume $b<0$, and note that $\al$ and $\bet$ are related through
\be
\al(x) = -\eta \, \bet(x)^{1/2} \, ,\label{BOU}
\ee 
where we have defined 
\be
\eta = \sqrt{\f{-b}{a\sqrt{3}}} \, ,
\ee
therefore the solution satisfies a mixed boundary condition.  Following the prescription in \cite{WITMULTI}, the boundary 
condition (\ref{BOU}) corresponds to deforming the boundary theory by 
\be
W=-\f{2\et }{3}\, \int d^3x\, {\cal O}_1(x)^{3/2}\, .\label{W}
\ee
This is so because $\lan {\cal O}_1(x) \ran = \bet(x)$, and hence (\ref{BOU}) can formally be written as
\be
\al(x) = \f{\del W}{\del \bet}\, .
\ee 
From (\ref{W}) we see that $\eta$ is the deformation parameter of the boundary theory, so that different values of $\eta$ define different 
boundary deformations. 
Hence, for a fix value of $\eta$, we are left with four moduli parameters for solution (\ref{SOL}). As we will see next, this matches the moduli 
parameters of the solution in the field theory side.

What are the corresponding dual boundary operators to the solution (\ref{SOL})? To answer this question let us consider 
$S^7$ as a $U(1)$ bundle over $CP^3$. Now we note that 
in the first ansatz, since spinors $\th$ are covariantly constant, they are invariant under $SU(4)$ 
isometry group of $CP^3$ but are charged under $U(1)$. This property is inherited by $L$ \cite{WARN}, and thus from our 
ansatz for $A_3$ we conclude that the dual operator should be invariant under $SU(4)$ global symmetry of the field 
theory. However, as the $U(1)$ isometry of the bulk is identified with the baryonic number symmetry $U(1)_b$, 
the dual operator is charged under $U(1)_b$ and so it comes with a monopole operator. 
For the second ansatz we note that $J$, the K\"ahler form, is invariant under $SU(4)$. It can also be shown that $e^7$ 
is invariant under this group. Further, $J$ and $e^7$ carry no charge under $U(1)$ so the whole ansatz is 
invariant under $SU(4)\times U(1)$. Therefore, the dual operator on the boundary must be a singlet under 
$SU(4)\times U(1)_b$. As already noted, the scaling behavior of the bulk solution near the boundary indicates that these operators 
must have dimension one or two. 

Let us see if we can identify such operators in the ABJM boundary theory. In this model a Chern-Simons-matter theory describes 
the boundary dynamics, where scalars $X^I=(Y^A, Y^{\dagg }_A)$ transform as 
\be
{\bf 8}_v= \bf{4} \oplus \bf {\bar{4}}
\ee
under $SU(4)$ R-symmetry. However, the only singlet scalar operator of dimension $1$ that one can construct is the Konishi operator, i.e. $\tr(Y^AY^\dagg _A)$, which as we know is not a BPS operator. Moreover, the Konishi operator 
is invariant under the whole $SO(8)$ symmetry group whereas our ansatz in the bulk is only invariant under $SU(4)$ subgroup. All these indicate that the dual operator cannot be the Konishi operator. 
For dual  operators of dimension two, we note that the fermionic fields are in 
${\bf 8}_c=\bf{4} \oplus \bf {\bar{4}}$, hence the second 
descendant of $\tr(X^{\{I}X^{J\}})$ (of dimension 2) contains ${\bf 35}_c={\bf 15} \oplus {\bf 10} \oplus {\bf 10}$ with no $SU(4)$ singlet.

Note that had we chosen the background field as in ABJM, i.e., (\ref{N2}), with the above 
ansatz we obtain scalars of $m^2=10$ (instead of getting $m^2=-2$ for skew-whiffed) which are singlets under $SU(4)$. These modes, however, can be recognized 
as the sixth descendant of $\tr(X^{\{I}X^JX^K X^{L\}})$. They sit in ${\bf 35}_s$ as indicated in table 1 
of \cite{PIO}, and this representation -- in the sixth descendant -- contains three $SU(4)$ singlets of $\Delta =5$  which we identify with the three scalars with $m^2=10$ in the ABJM background (\ref{N2}). However, if we assume the low energy dynamics of 
anti-membranes is also given by the ABJM (Chern-Simons-matter) theory we run into problem in identifying the dual operators. 

Recall that our solution had a flipped sign of $F_4$ and hence we should really discuss the theory of anti-membranes on 
the boundary. It has been observed that the spectrum of the Kaluza-Klein modes of the bulk supergravity of the two theories are related by interchanging the two representations ${\bf 8}_s$ and ${\bf 8}_c$ of $SO(8)$ \cite{HOP, KAL}. In ABJM theory supercharges and fermionic matter fields sit in ${\bf 8}_s$ and ${\bf 8}_c$, respectively. Therefore, for 
anti-membranes boundary theory we propose fermions to sit in ${\bf 8}_s$, which under $SU(4)$ decompose as 
\be
{\bf 8}_s= \bf{6} \oplus \bf {{1}} \oplus \bf{1}\, .\label{DEC}
\ee
Triality of $SO(8)$ then implies that the supercharges and scalars should decompose as 
\be
\bf{8}= \bf{4} \oplus \bf {\bar{4}}\, .
\ee

What is the field theory describing the low energy dynamics of anti-membranes? First note that the skew-whiffed bulk solution preserves supersymmetry only when $k=1$, i.e., when we have an $AdS_4\times S^7$ \cite{KAL} for which the  boundary theory should have an ${\cal N}=8$ supersymmetry. On the other hand, ABJM theory has been conjectured to have  an enhanced ${\cal N}=8$ supersymmetry for $k=1$ and $k=2$ \cite{ABJM}. This happens because of the existence of  appropriate monopole operators allowing to impose reality conditions on dynamical fields. The enhanced ${\cal N}=8$ supersymmetry of ABJM theory for $k=1$ and $k=2$ has further been investigated in \cite{REY, OH, WIM}. In particular, it 
has been shown that in these cases the action has a manifest $SO(8)$ symmetry. One can hence arrange the scalars, fermions, and the 
supercharges in three $SO(8)$ representations of ${\bf 8}_v$, ${\bf 8}_c$, and ${\bf 8}_s$, respectively.   

The anti-membranes action (for $k=1$) can therefore be obtained from ABJM action by a parity transformation and 
swapping ${\bf 8}_s$ and ${\bf 8}_c$ representations of $SO(8)$ by triality. The explicit form of the action can be worked out 
as in \cite{REY}, though, we do not need it as we will turn on only one $SU(4)$ singlet spinor and the gauge fields. Setting $Y^A$'s 
to zero all the interaction terms (including the fermionic ones) disappear so that we are left with the kinetic term of the spinor 
field and the Chern-Simons term. On the other hand, since we have chosen $\al$ in (\ref{AL}) to act as a source for the operator 
${\cal O}_1$ (of dimension $2$), turning on the scalar mode in the bulk corresponds to adding $W$ term (\ref{W}) to the boundary action:
\be
S\to S+ W\, .
\ee

Let us call $\ps$ the singlet spinor field in (\ref{DEC}) and see if the new action admits nontrivial classical solutions when we turn on only 
the $\ps$ field. By looking at the field equations, however, it is seen that we need also turn on the gauge fields. In this case, the $SU(4)$ 
singlet operator is 
\be
{\cal O}_1= \tr ({\bar \psi}\ps)\, ,
\ee 
and hence the deformed action, with only $\ps$ and the gauge fields turned on, becomes
\bea
S\!\!&=&\!\!\!\int d^3x\, \tr\!\lf( i{\bar \psi} {D\!\!\!\! /}\, \ps 
- \f{ik}{4\pi}\ep^{\mu\nu\rh}(A_\mu\pl_\nu A_\rh +\f{2i}{3} A_\mu A_\nu A_\rh 
-\Ah_\mu\pl_\nu \Ah_\rh -\f{2i}{3} \Ah_\mu \Ah_\nu \Ah_\rh ) \ri) \nn \\
&-&\! \f{2\eta}{3}\,  \int d^3x \lf(\tr ({\bar \psi}\ps)\ri)^{3/2}\, ,
\eea
where we have included an extra $i$ factor in front of the Chern-Simons term because of Euclidean signature.
The field equations then read
\bea
&&i{D\!\!\!\! /}\, \ps - \eta \, \lf(\tr ({\bar \psi}\ps)\ri)^{1/2}\ps =0 \, ,\nn \\
&&\ \ \ \ \ \f{ik}{4\pi}\ep^{\mu\nu\rh}F_{\nu\rh} +{\bar \ps}\ga^\mu \ps =0\, , \nn \\
&&\ \ \ \ \ \f{ik}{4\pi}\ep^{\mu\nu\rh}{\hat F}_{\nu\rh} +\, {\bar \ps}\ga^\mu \ps =0\, .\label{SEL}
\eea 

To find a solution we simply set
\be
\ps^a_\ah =\f{\del^a_\ah}{N}\, \ps
\ee
which is then equivalent to looking at the two $U(1)$ sectors of the gauge group. Now we can set the $SU(N)$ gauge 
fields to zero. For the $U(1)$ part, let $A^\pm_\mu = A_\mu \pm {\hat A}_\mu$, then we have
\bea
&&\f{ik}{4\pi}\ep^{\mu\nu\rh}F_{\nu\rh}^+ = -2{\bar \ps}\ga^\mu \ps \nn \\
&&\ \ \ \ \ {F}_{\nu\rh}^- = 0\, .\label{F}
\eea 
So we can further set $A^-_\mu =0$. However, note that the matter field only couples to $A^-_\mu$, so setting 
$A^-=0$ we are left with a self-interacting spinor field in the first equation of (\ref{SEL}). 
As the solutions we found in the bulk are nonsingular and spherically symmetric near the boundary, 
to solve the $\psi$ equation we make the following ansatz which is similarly nonsingular and rotationally symmetric:
\be
\ps= \f{(c +i ({x}-{x_0})^\mu \si_\mu)}{(c^2+(\vec{x}-\vec{x_0})^2)^{\ga}}\, \xi \, ,
\ee
with $c$ a free constant and $\xi$ an arbitrary constant spinor. This ansatz has been proposed earlier in 
solving the Seiberg-Witten equations on ${\bf R}^3$  \cite{NASH}, and interestingly the solution we obtain will  be 
identical to theirs up to a constant. 
Plugging this ansatz into the field equation of $\ps$, the normalization constant and $\ga$ get fixed
\be
\ps= \f{3c\sqrt{N}}{\eta}\, \f{(c +i ({x}-{x_0})^\mu\si_\mu)}{(c^2+(\vec{x}-\vec{x_0})^2)^{3/2}} \lf(
\begin{array}{l}
1 \\
0
\end{array}
\ri)\, .\label{S}
\ee

Further, let us compute the action of the above solution (with $A^-=0$):
\be
S=\int d^3x \lf[ \tr(i{\bar \psi} {\pl\!\!\! /}\, \ps) - \f{2\eta}{3}\, \lf(\tr ({\bar \psi}\ps)\ri)^{3/2}\ri]\, .
\ee 
Using the field equations and plugging (\ref{S}) into the action we obtain
\be
S= \f{\eta}{3}\, \int d^3 x \lf(\tr ({\bar \psi}\ps)\ri)^{3/2} = \f{9c^3}{\eta^2} \int \f{d^3x}{(c^2+(\vec{x}-\vec{x_0})^2)^3}
=\f{9\pi^2}{4\eta^2}\, . 
\ee      
Having a finite action, solution (\ref{S}) thus represents an instanton of the deformed boundary theory. 
For the gauge field $A^+_\mu$, we plug in solution (\ref{S}) into eq. (\ref{F}) and take the integral
\be
\oint_s F^+ = \oint_s \ep^{\mu\nu\rh}F^+_{\nu\rh}\, ds_\mu = 0\, ,
\ee
with $s$ a round sphere at infinity. Therefore this solution has no net magnetic charge, and we can consistently 
identify it with the solution in the bulk which is invariant under $U(1)$ isometry group ($SU(4)\times U(1) \subset SO(8)$).

One can also examine the correlation functions of $\ps$'s in instanton background (\ref{S}). In particular, 
we can obtain the dominant contribution to the expectation value of ${\cal O}=\tr ({\bar \psi}\ps)$ by evaluating 
it in this background   
\be
\tr ({\bar \psi}\ps)= \f{9c^2}{\eta^2\, (c^2+(\vec{x}-\vec{x_0})^2)^2}\, . 
\ee 
Moreover, if we set $c^2 =a^2-b^2$, this will be proportional to (\ref{BET}), the expectation value we obtained in the 
bulk by analyzing the behavior of solution (\ref{SOL}) near the boundary. So, as expected, the field theory analysis 
is consistent with the bulk computations. Also, note that the moduli parameters of the bulk and boundary solutions match.

\section{Conclusions and Outlook}
In this paper we provided two ansatzs which reduced the 11d  supergravity field equations on the 
background of skew-whiffed $AdS_4\times S^7$ to a 
4d conformally coupled scalar equation. We found the exact solutions and examined their behavior near the 
boundary. The scalar modes turned out to be singlets under $SU(4)$ subgroup of $SO(8)$ with $m^2=-2$. 
Our main task was to find the dual operators to these modes. We argued that there are no BPS operators in 
the boundary ABJM theory whose quantum numbers could match with those of bulk scalars. Therefore, we inferred 
that the boundary theory of anti-membranes cannot be identical with the theory of membranes. 
A crucial hint came from the skew-whiffed bulk theory where one has to swap the $s$ and $c$ representations 
for consistency. Hence we proposed the theory of anti-membranes should analogously be obtained from ABJM by 
interchanging the $s$ and $c$ representations. Doing so, we were able to identify the BPS operators corresponding 
to the scalar modes that we found in the bulk. On the field theory side, we deformed the action by the dual operator 
and found an exact classical solution which we identified with the bulk solution invariant under $U(1)\times SU(4)$. 
Apart from this, there are two more operators (being complex conjugate of each other) that are invariant under 
$SU(4)$ but carry $U(1)$ charge, and so contain monopole operators. These correspond to bulk solutions that we found in our first ansatz. 

Our analysis of anti-membranes theory provides a realization of the boundary toy model discussed in \cite{DEH, PAPA}. Therefore, it 
is interesting to study the instability of the bulk vacuum through the instantons discussed above. Similarly, the instabilities of 
the supergravity solutions that have recently been studied in \cite{BOB} should be understandable in this framework.

\vspace{3mm}

\hspace{-6mm}{\large \textbf{Acknowledgments}}

\vspace{1.5mm}

\noindent
A.I. acknowledges the support by the University of Tarbiat Modares during his sabbatical leave.  
He would also like to thank P. Bouwknegt for support and hospitality while he was visiting the Department of theoretical physics at the Australian National University, where part of this work was done. He is grateful to 
M. Staudacher for useful discussions. M.N. thanks M. Bianchi for discussions and hospitality during his visit to the Department of physics at the  University of Rome II.

\end{document}